\documentclass[aip,apl,amsmath,amssymb,reprint,superscriptaddress,longbibliography]{revtex4-1}
\usepackage{graphicx}% Include figure files
\usepackage{color}
\begin{document}
\title{Detection and control of single proton spins in a thin layer of diamond grown by chemical vapor deposition}
\author{Kento Sasaki}
\affiliation{School of Fundamental Science and Technology, Keio University, 3-14-1 Hiyoshi, Kohoku-ku, Yokohama 223-8522, Japan}
\author{Hideyuki Watanabe}
\affiliation{Device Technology Research Institute, National Institute of Advanced Industrial Science and Technology,
Tsukuba Central 2, 1-1-1 Umezono, Tsukuba, Ibaraki 305-8568, Japan}
\author{Hitoshi Sumiya}
\affiliation{Advanced Materials Laboratory, Sumitomo Electric Industries, Ltd., Itami, Hyogo 664-0016, Japan}
\author{Kohei M. Itoh}
\email{kitoh@appi.keio.ac.jp}
\affiliation{School of Fundamental Science and Technology, Keio University, 3-14-1 Hiyoshi, Kohoku-ku, Yokohama 223-8522, Japan}
\affiliation{\mbox{Center for Spintronics Research Network, Keio University, 3-14-1 Hiyoshi, Kohoku-ku, Yokohama 223-8522, Japan}}
\author{Eisuke Abe}
\email[]{Author to whom correspondence should be addressed: eisuke.abe@riken.jp}
\affiliation{School of Fundamental Science and Technology, Keio University, 3-14-1 Hiyoshi, Kohoku-ku, Yokohama 223-8522, Japan}
\affiliation{RIKEN Center for Emergent Matter Science, Wako, Saitama 351-0198, Japan}
\date{\today}
\begin{abstract}
We report detection and coherent control of a single proton nuclear spin using an electronic spin of the nitrogen-vacancy (NV) center in diamond as a quantum sensor.
In addition to determining the NV--proton hyperfine parameters by employing multipulse sequences,
we polarize and coherently rotate the single proton spin, and detect an induced free precession.
Observation of free induction decays is an essential ingredient for high resolution proton nuclear magnetic resonance, and the present work extends it to the atomic scale.
We also discuss the origin of the proton as incorporation during chemical vapor deposition growth,
which provides an opportunity to use protons in diamond as built-in quantum memories coupled with the NV center.
\end{abstract}
\maketitle
Hydrogen $^{1}$H atoms are ubiquitous in organic compounds and human body.
Proton nuclear magnetic resonance (NMR), including magnetic resonance imaging (MRI), has established itself as
an essential tool for molecular structure analysis and clinical diagnosis.
Reducing the sample volume down to the atomic scale is tantalizing technological advance that potentially innovates these fields.
A leading candidate to achieve this feat, currently under intensive research,
is a quantum sensor based on a single nitrogen-vacancy (NV) center in diamond.~\cite{MKS+13,SSP+13,MKC+14,HSR+15,DPL+15,APN+17}
Indeed, detection of single protons, as well as spectroscopy of single proteins, has been demonstrated using the NV sensor,
although the number of reports is still limited.~\cite{LSU+16,SLC+14}
The present work reports detection, as well as coherent control, of elusive single protons via the NV center.
We observe free induction decays from a single proton,
bringing the NV-based atomic scale NMR closer to the setting of existing high resolution NMR spectroscopy.
Single protons we detect are embedded in a thin layer of diamond grown by microwave plasma assisted chemical vapor deposition (CVD).
This has further implications in quantum information processing and materials science;
single proton spins in diamond potentially function as quantum memories networking with the NV center qubit,~\cite{MKL+12,BRA+19}
whereas the process of hydrogen incorporation into diamond during the CVD growth is hardly understood.

The diamond sample we examined is a thin layer of diamond CVD-grown on a type-IIa (100) single crystal diamond prepared
by the high pressure--high temperature (HPHT) process.
The isotopic composition of the substrate is natural abundance ($^{12}$C : $^{13}$C = 98.9\% : 1.1\%) with trace amount of nitrogen ($<$ 0.1~ppm).~\cite{SS96}
Throughout the CVD growth, the microwave power, the chamber pressure,
and the substrate temperature were kept at 750~W, 25~Torr, and 800~$^{\circ}$C, respectively.~\cite{WTY+99}
The feed gas was a mixture of hydrogen (H$_2$) and methane (CH$_4$) with isotopically purified $^{12}$C (99.999\%).~\cite{IW14}
First, an undoped buffer layer was grown under the condition of [CH$_4$]/[H$_2$] = 0.025\%.
The doped layer of a few tens of nanometers was then grown under condition of [CH$_4$]/[H$_2$] = 0.5\%, with the nitrogen-to-carbon ratio of 24\%.
As a side note, the natural abundance of nitrogen is $^{14}$N : $^{15}$N = 99.63\% : 0.37\%.
We thus can discriminate the NV centers in the CVD layer from those in the substrate unambiguously,
by observing an isotope-dependent hyperfine splitting of the $^{15/14}$N nuclear spin ($I$ = $\frac{1}{2}$ or 1) intrinsic to the NV center.~\cite{AS18}
In reality, we never found $^{14}$NV from the present sample.

Our measurement setup is a homebuilt confocal microscope
using a 515-nm excitation laser and a single-photon counting module for optically initializing and detecting the electronic spin of the NV center.~\cite{AS18,SIA18}
A copper wire runs across the diamond surface, delivering microwave and radiofrequency (rf) magnetic fields
to control the NV electronic spin and the proton nuclear spin, respectively.
The $m_S$ = 0 and $-$1 states of the $S$ = 1 NV spin were used for quantum sensing, and the microwave frequency was tuned to the corresponding transition. 
A position-controlled permanent magnet was used to provide a static magnetic field $B_0$ of
%\color{red}
5--45~mT,
%\color{black}
which was applied parallel to the NV symmetry axis.
All the measurements were performed under ambient conditions.

\begin{figure}
\begin{center}
\includegraphics{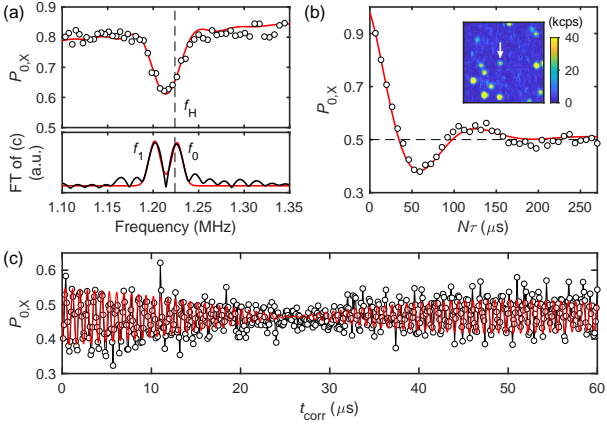}
\caption{Detection of a single proton nuclear spin.
(a) NMR spectrum at 28.7~mT (top panel).
The solid line is a simulation using the hyperfine constants determined from (b) and (c).
(b) Coherent driving of the proton nuclear spin by increasing the number of pulses $N$.
The solid line is a fit with a damped sinusoidal function.
The inset shows a fluorescence image of the sample surface (10$\times$10~$\mu$m$^2$).
The arrow indicates the position of NV1.
(c) Correlation spectroscopy.
The solid line is a fit with a two-component sinusoidal oscillation.
Fourier spectrum is shown in the bottom panel of (a).
\label{fig1}}
\end{center}
\end{figure}

In what follows, we first present a series of data obtained from one particular NV center [labeled as NV1. see the inset of Fig.~\ref{fig1}(b) for a fluorescence image],
which we will reveal is coupled with a single proton.
The data from other NV centers and the discussion on the origin of the proton will be given later.

NMR spectra were recorded by multipulse sequences that periodically invert the NV spin by applying microwave $\pi$ pulses every $\tau$ seconds.
More specifically, we applied XY16-$N$ sequences, in which the XY16 sequence, given by
$\tau$/2--X--$\tau$--Y--$\tau$--X--$\tau$--Y--$\tau$--Y--$\tau$--X--$\tau$--Y--$\tau$--X--$\tau$--\={X}--$\tau$--\={Y}--$\tau$--\={X}--$\tau$--\={Y}--$\tau$--\={Y}--$\tau$--\={X}--$\tau$--\={Y}--$\tau$--\={X}--$\tau$/2,
is repeated $N$/16 times.~\cite{GBC90}
X, Y, \={X}, and \={Y} denote the $\pi$ pulses about $x$, $y$, $-x$, and $-y$ axes of the rotating frame, respectively.
The full sequence is given by
L--X/2--(XY16-$N$)--X/2--L$_{\mathrm{RO}}$,
where L denotes a laser illumination, the suffix RO refers to photon counting during L, and X/2 denotes a $\pi$/2 pulse about the $x$ axis.
In each run, the interpluse delay $\tau$ is fixed, and is swept for different runs.
When $\tau$ coincides with half the Larmor period of a target nuclear spin, the decoherence of the NV spin is enhanced.
The circle points ($\bigcirc$) in Fig.~\ref{fig1}(a) show an NMR spectrum as a function of $(2\tau)^{-1}$ taken by XY16-64.
$B_0$ was 28.7~mT, which sets the proton Larmor frequency $f_{\mathrm{H}} = \gamma_{\mathrm{H}} B_0/2\pi$ at 
%\color{red}
1.2239~MHz,
%\color{black}
with $\gamma_{\mathrm{H}}/2\pi$ = 42.577~kHz/mT the gyromagnetic ratio of $^{1}$H.
The vertical axis is labeled as $P_{0,\mathrm{X}}$.
$P_0$ is the probability of the NV spin being in $m_S$ = 0, and the suffix X specifies the axis of the readout pulse.
%\color{red}
The NMR resonance is marked by the decrease in $P_{0,\mathrm{X}}$.
As observed in Fig.~\ref{fig1}(a), the minimum of the dip $f_{\mathrm{XY}}$ does not match with $f_{\mathrm{H}}$.
%\color{black}
This hints the presence of the hyperfine interaction with the NV center,
which, if only one nucleus is coupled, will be characterized by the parallel and perpendicular components $A_{\parallel}$ and $A_{\perp}$.

The NV--proton coherent coupling can be verified by increasing the number of pulses $N$ (up to 656),
with $\tau$ fixed at 411.5~ns = (2$\times$1.2151~MHz)$^{-1}$ [Fig.~\ref{fig1}(b)].
By detecting the proton spin stroboscopically only at the end of the $N$-pulse sequences,
the proton spin is seen to rotate about the $A_{\perp}$ axis, when $B_0 \gg A_{\parallel,\perp} (2\pi/\gamma_{\mathrm{H}}$).~\cite{KUBL12,TWS+12,BCA+16,MSI+20}
The NMR signal oscillates at the frequency $f_{\mathrm{osc}}$ of 7.414~kHz.
%\color{green}
Crucially, the quantum nature of the coupling is evident from the data in which $P_{0,\mathrm{X}}$ goes well below 0.5.~\cite{KUBL12,TWS+12,ZHH+11,ZHS+12}

Correlation spectroscopy enables a further analysis of the NV--proton interaction.~\cite{LDB+13,KSD+15,SRP+15}
%\color{black}
The sequence is given by
L--X/2--(XY16-32)--Y/2--$t_{\mathrm{corr}}$--Y/2--(XY16-32)--X/2--L$_{\mathrm{RO}}$,
where $t_{\mathrm{corr}}$ is swept, with $\tau$ fixed at 412.5~ns = (2$\times$1.2121~MHz)$^{-1}$. 
Figure~\ref{fig1}(c) demonstrates a two-component oscillation with frequencies at $f_0$ = 1.2234~MHz and $f_1$ = 1.2046~MHz [the bottom panel of Fig.~\ref{fig1}(a)].
$f_0$, matching well with $f_{\mathrm{H}}$, arises from the coupling to $m_S$ = 0, whereas the coupling to $m_S$ = $-$1 shifts the proton precession frequency to $f_1$.
The average of the two, $(f_0 + f_1)/2$ = 1.2140~MHz, corresponds to the NMR resonance
%\color{red}
frequency $f_{\mathrm{XY}}$ obtained above.
%\color{black}
From analytic formulae of $A_{\parallel}$ and $A_{\perp}$ expressed in terms of $f_{\mathrm{osc},0,1}$ and $\tau$ [used in Fig.~\ref{fig1}(b)],~\cite{TWS+12,BCA+16,SIA18}
we obtain $A_{\parallel}/2\pi$ = $-$19.0~kHz and $A_{\perp}/2\pi$ = 22.9~ kHz.
These values are fed back to the simulation of the NMR spectrum,
and reproduce the experimental data excellently, as shown by the solid line in Fig.~\ref{fig1}(a).
The consistent analysis here strongly supports the observation of a single proton weakly coupled to a single NV center.

\begin{figure}
\begin{center}
\includegraphics{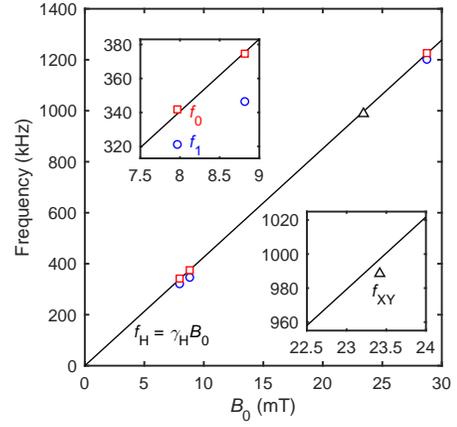}
\caption{
%\color{blue}
$B_0$-dependence of the signal frequencies.
The solid line is given as $f_{\mathrm{H}} = \gamma_{\mathrm{H}} B_0$, and is not a fit.
The points at the largest $B_0$ correspond to the data in Fig.~\ref{fig1}(c).
The insets are close-ups around 8~mT and 23.5~mT.
The error bars are within the sizes of the data symbols in both horizontal and vertical directions.
%\color{black}
\label{fig2}}
\end{center}
\end{figure}

%\color{blue}
In Fig.~\ref{fig2}, we examine $B_0$-dependence of the signal frequencies.
$f_0$ ($\square$) closely follows $f_{\mathrm{H}}$,
whereas $f_1$ ($\bigcirc$) and $f_{\mathrm{XY}}$ ($\triangle$) are smaller than $f_{\mathrm{H}}$, as expected from the analysis above.
We note that correlation spectroscopy is free from spurious harmonics,
excluding the possibility that the signals are associated with $^{13}$C nuclei.~\cite{LBR+15,BCA+16}
Spurious harmonics of $^{13}$C nuclei are also evaded by the use of spinless $^{12}$C.
Furthermore, undersampling of $f_1$, which can arise when $|A_{\parallel}|$ is large compared with the sampling rate of correlation spectroscopy (10~MHz here),
is also excluded by the fact that $f_0 - f_1$ is almost constant.
%\color{black}

If a purely dipolar coupling is assumed, the coupling constants are explicitly written as
$A_{\parallel} = (\hbar \gamma_{\mathrm{e}} \gamma_{\mathrm{H}} \mu_0/4\pi) \times (3 \cos^2 \theta -1)/r^3$ and
$A_{\perp} = (\hbar \gamma_{\mathrm{e}} \gamma_{\mathrm{H}} \mu_0/4\pi) \times 3 \sin \cos \theta/r^3$,
with $\hbar$ the reduced Planck constant, $\gamma_{\mathrm{e}}$/2$\pi$ = 28.0~MHz/mT the gyromagnetic ratio of the NV spin,
$\mu_0$ the permeability of the vacuum, $\theta$ the polar angle of the proton relative to the NV axis, and $r$ the NV--proton distance.
These relations give estimates of the spatial coordinates as $\theta$ = 72.3$^{\circ}$ and $r$ = 1.44~nm,
providing a direct route to the atomic scale NMR.~\cite{SIA18,BCA+16,ZCS+18,ZHCD18,ARB+19}

\begin{figure}
\begin{center}
\includegraphics{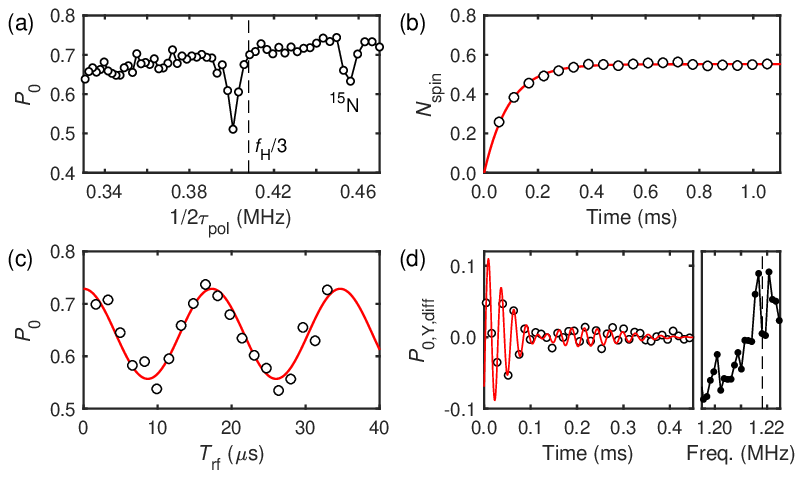}
\caption{Control of a single proton nuclear spin.
(a) PulsePol spectrum around one-third of the proton Larmor frequency.
(b) Transient evolution of polarization transfer.
The solid line is an exponential fit.
(c) Rabi driving.
The solid line is a fit with a sinusoidal function.
(d) Free induction decay (left), and its Fourier spectrum (right).
The solid line is a fit with a two-component sinusoidal damped oscillation.
The dashed line indicates $f_{\mathrm{p}}$ (see the main text for detail).
\label{fig3}}
\end{center}
\end{figure}

Having identified and characterized a single proton, we now demonstrate the polarization, coherent rotation, free precession of the single proton nucleus.
Pulsed dynamic nuclear polarization (PulsePol) is a technique that transfers an electron spin polarization to nuclear spins
through dynamical control of the interaction between the electron and nuclear spins by microwave pulses.~\cite{SST+18}
We make use of two sequences which we call PolY and PolX.~\cite{SIA18}
The PolY sequence is given by
Y/2--$\tau_{\mathrm{pol}}$/4--\={X}--$\tau_{\mathrm{pol}}$/4--Y/2--X/2--$\tau_{\mathrm{pol}}$/4--Y--$\tau_{\mathrm{pol}}$/4--X/2--Y/2--$\tau_{\mathrm{pol}}$/4--\={X}--$\tau_{\mathrm{pol}}$/4--Y/2--X/2--$\tau_{\mathrm{pol}}$/4--Y--$\tau_{\mathrm{pol}}$/4--X/2,
where $\tau_{\mathrm{pol}}$ is the interpluse delay.
PolX is PolY in the reverse order, and PolY and PolX have an opposite effect on the sign of the polarization transfer.~\cite{SIA18}
The total duration (2$\tau_{\mathrm{pol}}$) is adjusted close to a commensurate condition $2\tau_{\mathrm{pol}} = 3/f_{\mathrm{H}}$,
where the polarization transfer is expected to occur.
%\color{red}
Figure~\ref{fig3}(a)
%\color{black}
shows a PulsePol spectrum as sweeping 2$\tau_{\mathrm{pol}}$.
Each data point was taken by the sequence
(PolX)$^{20}$--L--(PolY)$^{20}$--L$_{\mathrm{RO}}$, where the superscript specifies the number of repetitions. 
The first (PolX)$^{20}$ is to cancel out any nuclear polarization during a preceding run.
The NV polarization is recovered by L, and transferred to the nucleus by (PolY)$^{20}$.
The decrease in the NV polarization is detected during the second L, which also re-initializes the NV spin for the next run.
The spectrum reveals the first dip near $f_{\mathrm{H}}/3$ = 0.4080~MHz (the dashed line) and the second one at 0.4562~MHz.
The exact position of the first dip is 0.4006~MHz, which is closer to the one-third of the NMR dip in Fig.~\ref{fig1}(a).
The second dip has been analyzed as arising from $^{15}$N nucleus of the NV center itself, and is not of central interest here.

The degree of polarization transfer can be evaluated quantitatively by the sequence
[(PolX)$^{20}$--L]$^{20}$--[(PolY)$^{20}$--L$_{\mathrm{RO}}$]$^{20}$,
where 2$\tau_{\mathrm{pol}}$ in PolX/Y is fixed at 2.4960~$\mu$s = 3/1.2019~MHz.~\cite{SSM+17}
As (PolY)$^{20}$--L$_{\mathrm{RO}}$ is repeated, $P_0(n)$, which is $P_0$ after the $n$th (PolY)$^{20}$, evolves from $P_0(1)$ = 0.48
to the saturation value $P_{0,\mathrm{sat}}$ = 0.73, already reached at $n \approx$ 6.
$P_{0,\mathrm{sat}}$ is primarily determined by the decoherence of the NV spin during the sequence.
$P_{0,\mathrm{sat}} - P_0(n)$ quantifies the amount of polarization (in units of $\hbar$) transferred to the nuclear spin by the $n$th (PolY)$^{20}$.
Therefore, $P_0(n)$ normalized by the integration of $P_{0,\mathrm{sat}} - P_0(n)$ from $n$ = 1 to 6
gives the cumulative number of inverted nuclear spins $N_{\mathrm{spin}}$, which is plotted in
%\color{red}
Fig.~\ref{fig3}(b).
%\color{black}
The solid line is a fit given by an empirical form
$N_{\mathrm{spin}} = N_{\mathrm{spin,sat}}(1-e^{-t/t_{\mathrm{c}}}$),
with $t$ the evolution time and $t_{\mathrm{c}}$ the time constant.
We find $N_{\mathrm{spin}}$ saturate at $N_{\mathrm{spin,sat}}$ = 0.56, with $t_{\mathrm{c}}$ = 93~$\mu$s.
While $N_{\mathrm{spin}}$ should reach one ideally in the case of a single nuclear spin,
the observed $N_{\mathrm{spin}} < 1$ provides a further evidence of the presence of a single proton.
A possible reason of $N_{\mathrm{spin}}$ not reaching unity is that the polarization rate competes with the relaxation rate of the nuclear spin.

The polarized nucleus can be driven by an rf pulse.
%\color{red}
Figure~\ref{fig3}(c)
%\color{black}
demonstrates a nuclear Rabi oscillation recorded by the sequence
[(PolY)$^{20}$--L]$^{10}$--$T_{\mathrm{rf}}$--[(PolY)$^{20}$--L$_{\mathrm{RO}}$]$^{10}$,
where $T_{\mathrm{rf}}$ is the length of an rf pulse with its frequency set at 1.2151~MHz.
The (PolY)$^{20}$--L block is repeated 20 times during one cycle, so that the nuclear spin is always polarized prior to an rf pulse.
In plotting
%\color{red}
Fig.~\ref{fig3}(c),
%\color{black}
only the first RO data of respective $T_{\mathrm{rf}}$ were used (other RO data were to monitor the re-polarization process).
The sinusoidal fit determines the Rabi frequency to be 57.7~kHz, setting the length of the $\pi$/2 pulse as $T_{\mathrm{rf,\pi/2}}$ = 4.115~$\mu$s.
The rf $\pi$/2 pulse induces a free precession of the polarized nuclear spin, which is viewed as a phase-coherent oscillation by the NV center and is measured by the sequence 
[(PolY/X)$^{20}$--L]$^{5}$--$T_{\mathrm{rf,\pi/2}}$--[X/2--(XY16-16)--Y/2--L$_{\mathrm{RO}}$]$^{50}$.
The choice of PolY or PolX inverts the sign of the resulting oscillation, as the nuclear spin is polarized to the opposite direction.~\cite{SIA18}
The detection part after $T_{\mathrm{rf,\pi/2}}$ is repeated 50 times in series, with $t_{\mathrm{s}}$ = $N\tau$ in XY16-16 set as 16$\times$411.5~ns = 6.584~$\mu$s.
X/2 pulses in adjacent blocks are separated by $t_{\mathrm{L}}$ = 11.840~$\mu$s.
In other words, a nuclear spin precession is continuously read out at the ``sampling frequency'' of 84.46~kHz (= 1/$t_{\mathrm{L}}$), undersampled at the 28th order.
The readout pulse is Y/2, so that the detection is phase-sensitive:
a common feature of high-resolution spectroscopy methods developed recently by several groups.~\cite{SGS+17,BCZD17,GBL+18}
During the continuous readout, the NV spin is either in a superposition of the $m_S$ = 0 and $-1$ states (between X/2 and Y/2, for the duration $t_{\mathrm{s}}$)
or in the $m_S$ = 0 state (otherwise, $t_{\mathrm{L}} - t_{\mathrm{s}}$).
Therefore, the averaged precession frequency experienced by the NV spin is expected to be
$f_{\mathrm{p}}$ = $(f_0 + f_1)/2 \times (t_{\mathrm{s}}/t_{\mathrm{L}}) + f_0 \times (t_{\mathrm{L}} - t_{\mathrm{s}})/t_{\mathrm{L}}$ = 1.2182~MHz.~\cite{PWS+19,CBH+19}
%\color{red}
Figure~\ref{fig3}(d)
%\color{black}
shows a free induction decay of a single proton nuclear spin thus obtained.
The difference between the signals by PolY and PolX are plotted (denoted as $P_{0,\mathrm{Y,diff}}$), and its Fourier spectrum takes into account the undersampling.
Remarkably, two peaks appears at $f_{\mathrm{a}}$ = 1.2173~MHz and $f_{\mathrm{b}}$ = 1.2203~MHz, on both sides of $f_{\mathrm{p}}$ (the dashed line).
This demonstrates the power of high resolution spectroscopy, which can reveal a fine structure that cannot be done otherwise.
The cause of the splitting is discussed below, in conjunction with the origin of the proton.

\begin{figure}
\begin{center}
\includegraphics{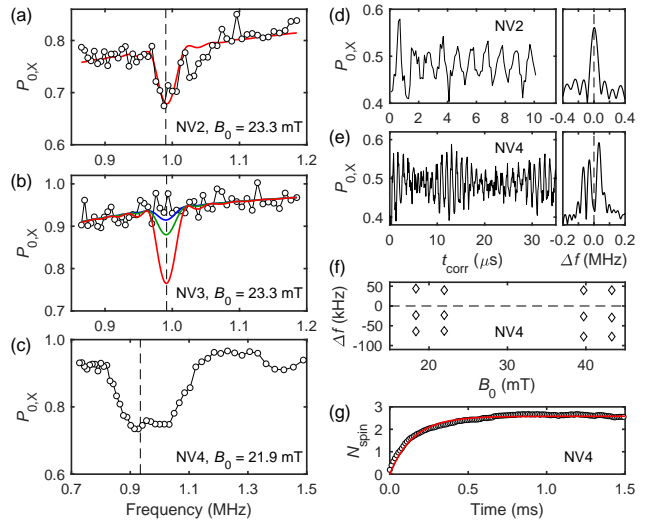}
\caption{
%\color{red}
(a--c) NMR spectra of (a) NV2 (ensemble, XY16-64), (b) NV3 (no signal, XY16-64), and (c) NV4 (single protons, XY16-16).
The dashed lines indicate the proton Larmor frequencies $f_{\mathrm{H}}$.
The solid line in (a) is a fit with $d_{\mathrm{NV}}$ = 10.3~nm.
The solid lines in (b) are simulations with $d_{\mathrm{NV}}$ = 10 (red), 15 (green), and 20~nm (blue).
(d--e) Correlation spectroscopy of (d) NV2 and (e) NV4.
Fourier spectra are shown as offsets from the proton Larmor frequencies ($\Delta f = f - f_{\mathrm{H}}$).
%\color{black}
%\color{magenta}
(f) The signal frequencies of NV4 as a function of $B_0$.
(g) Polarization transfer in NV4.
%\color{black}
\label{fig4}}
\end{center}
\end{figure}

At this point, we summarize the property of our sample.
From fluorescence images of the sample surface [e.g., the inset of Fig.~\ref{fig1}(b)],
the areal density of the NV centers is estimated to be 3$\times$10$^{6}$~cm$^{-2}$.
The fluorescence was obtained only at the surface (far below the depth resolution of our confocal microscope).
The most reliable approach to determining the depths of the NV centers in this case is to measure NMR of ensemble protons in immersion oil.~\cite{PDC+16}
In fact, all the measurements, including those for NV1, were performed with an oil immersion objective lens.
Out of 25 NV centers measured, 4 showed ensemble proton NMR, 19 no signals, 2 single protons.
Typical spectra of the respective cases (labeled as NV2, NV3, and NV4, respectively) are shown in
%\color{red}
Figs.~\ref{fig4}(a--c).
%\color{black}
The NMR spectrum of NV2 is fitted by assuming ensemble protons
%\color{red}
[Fig.~\ref{fig4}(a)],
%\color{black}
and the depth of the NV center ($d_{\mathrm{NV}}$) is estimated to be 10.3~nm.
See Ref.~\onlinecite{PDC+16} for detail of the fitting.
Correlation spectroscopy confirms a single-component oscillation at the proton Larmor frequency
%\color{red}
[Fig.~\ref{fig4}(d)].
%\color{black}
NV3 showed no signal
%\color{red}
[Fig.~\ref{fig4}(b)].
%\color{black}
The solid lines are calculated proton ensemble NMR spectra assuming $d_{\mathrm{NV}}$ = 10, 15 and 20~nm, suggesting that NV3 is located deeper than 20~nm.

%\color{magenta}
NV4 exhibited a broad NMR spectrum that is incompatible with proton ensemble in immersion oil [Fig.~\ref{fig4}(c)].
Correlation spectroscopy reveals that, unlike NV1, the oscillation is composed of more than two frequencies [Fig.~\ref{fig4}(e)].
Even more surprising is that the Fourier peaks do not appear at $f_{\mathrm{H}}$ ($\Delta f$ = 0).
On the other hand, $B_0$-dependence of the signal frequencies shows that they follow $f_{\mathrm{H}}$ [Fig.~\ref{fig4}(f)],
confirming that they are associated with protons.
We also perform polarization transfer similar to Fig.~\ref{fig3}(b), and observe that $N_{\mathrm{spin}}$ saturates at 2.6 [Fig.~\ref{fig4}(g)].
$N_{\mathrm{spin}} > 1$ evidences the presence of multiple protons, possibly three.

We interpret the cause of the shifts as due to the nuclear--nuclear interactions.
Similar observations have been made recently by Abobeih {\it et al.} in a cluster of $^{13}$C nuclei in diamond.~\cite{ARB+19}
They analyzed the nuclear--nuclear interactions by applying rf pulses and selectively turning on and off the interactions,
and imaged three-dimensional structure of a 27-nuclear-spin system.
The detailed analysis of the nuclear--nuclear and NV--nuclear interactions present in our system is beyond the scope of the present work,
but is an intriguing subject of future work.
The system may serve as a testbed for nanoscale MRI or multi-nuclear quantum memory.~\cite{BRA+19,ARB+19}
%\color{black}

%\color{cyan}
Lastly, we return to NV1 and discuss the origin of the proton.
The fact that the signal from NV1 is fully explainable from the hyperfine parameters, and that the signal from oil is absent, indicates that it is a deep NV center.
Accordingly, the observed proton, with the estimated NV1--proton distance of $r$ = 1.44~nm, cannot be interpreted as arising from the surface.
For instance, the surface of as-grown CVD diamond is terminated by hydrogen.
But if the signal of NV1 had arisen from the H-terminated surface, it would have overlapped with that of oil.
We thus conclude that protons were incorporated into diamond during CVD growth.

The form of hydrogen within the diamond matrix remains unidentified, but the splitting observed in Fig.~\ref{fig3}(d) provides a clue.
We speculate it as due to dipolar coupling between $^{1}$H and nearby $^{15}$N ($I$ = $\frac{1}{2}$).
If we assume the $^{1}$H--$^{15}$N distance to be on the order of the bond length of diamond ($r_{\mathrm{d}}$ = 0.154~nm),
the coupling strength is roughly estimated as $h \gamma_{\mathrm{N}} \gamma_{\mathrm{H}} \mu_0/4\pi r_{\mathrm{d}}^3$ = 3.33~kHz,
close to the observation ($\gamma_{\mathrm{N}}$: gyromagnetic ratio of $^{15}$N).
Complex defects containing both N and H, such neutral NVH ($S$ = 0), may exhibit such a coupling.~\cite{GNM+03,GBJS03,EDC+12}
%\color{black}

In conclusion, we detected, characterized, and polarized a single proton nuclear spin, and observed its free precession induced by an rf pulse via the NV center.
The spatial coordinates $r$ and $\theta$ were determined,
and the present setup is fully compatible with a protocol to determine the remaining parameter, the azimuthal angle $\phi$,
as demonstrated by some of the present authors and others recently.~\cite{SIA18,ZHCD18}
Observation of free induction decays is an essential ingredient that makes conventional high resolution proton NMR so powerful.
We extend it to the single spin level, and in fact revealed a fine structure in the proton spectrum.
The present work thus provides a route toward the atomic scale imaging by employing the state-of-the-art, but rather simple,
quantum sensing protocols, which so far have been tested primarily on $^{13}$C nuclei in diamond.~\cite{AS18,SIA18,ZCS+18,ZHCD18,ARB+19,PWS+19,CBH+19,DRC17}.
The protons we detected and controlled are interpreted to reside in diamond, presumably incorporated during CVD.
This points to the possibility to utilize protons in diamond as built-in quantum memories coupled with the NV center.
Compared with more common $^{13}$C, the higher Larmor frequency of the proton allows for faster operations.

K.S. was supported by the JSPS Grant-in-Aid for Research Fellowship for Young Scientists (DC1), Grant No.~JP17J05890.
H.W. was supported by the JSPS Grant-in-Aid for Scientific Research (KAKENHI) (B) Grant No.~18H01502.
K.M.I. was supported by the JSPS KAKENHI (S) Grant No.~26220602 and (B) Grant No.~19H02547,
the JST Development of Systems and Technologies for Advanced Measurement and Analysis (SENTAN),
and the Spintronics Research Network of Japan (Spin-RNJ).
The data that support the findings of the present work are available from the corresponding author upon reasonable request.
\bibliography{proton}

%merlin.mbs aipnum4-1.bst 2010-07-25 4.21a (PWD, AO, DPC) hacked
%Control: key (0)
%Control: author (8) initials jnrlst
%Control: editor formatted (1) identically to author
%Control: production of article title (0) allowed
%Control: page (1) range
%Control: year (1) truncated
%Control: production of eprint (0) enabled
\begin{thebibliography}{41}%
\makeatletter
\providecommand \@ifxundefined [1]{%
 \@ifx{#1\undefined}
}%
\providecommand \@ifnum [1]{%
 \ifnum #1\expandafter \@firstoftwo
 \else \expandafter \@secondoftwo
 \fi
}%
\providecommand \@ifx [1]{%
 \ifx #1\expandafter \@firstoftwo
 \else \expandafter \@secondoftwo
 \fi
}%
\providecommand \natexlab [1]{#1}%
\providecommand \enquote  [1]{``#1''}%
\providecommand \bibnamefont  [1]{#1}%
\providecommand \bibfnamefont [1]{#1}%
\providecommand \citenamefont [1]{#1}%
\providecommand \href@noop [0]{\@secondoftwo}%
\providecommand \href [0]{\begingroup \@sanitize@url \@href}%
\providecommand \@href[1]{\@@startlink{#1}\@@href}%
\providecommand \@@href[1]{\endgroup#1\@@endlink}%
\providecommand \@sanitize@url [0]{\catcode `\\12\catcode `\$12\catcode
  `\&12\catcode `\#12\catcode `\^12\catcode `\_12\catcode `\%12\relax}%
\providecommand \@@startlink[1]{}%
\providecommand \@@endlink[0]{}%
\providecommand \url  [0]{\begingroup\@sanitize@url \@url }%
\providecommand \@url [1]{\endgroup\@href {#1}{\urlprefix }}%
\providecommand \urlprefix  [0]{URL }%
\providecommand \Eprint [0]{\href }%
\providecommand \doibase [0]{http://dx.doi.org/}%
\providecommand \selectlanguage [0]{\@gobble}%
\providecommand \bibinfo  [0]{\@secondoftwo}%
\providecommand \bibfield  [0]{\@secondoftwo}%
\providecommand \translation [1]{[#1]}%
\providecommand \BibitemOpen [0]{}%
\providecommand \bibitemStop [0]{}%
\providecommand \bibitemNoStop [0]{.\EOS\space}%
\providecommand \EOS [0]{\spacefactor3000\relax}%
\providecommand \BibitemShut  [1]{\csname bibitem#1\endcsname}%
\let\auto@bib@innerbib\@empty
%</preamble>
\bibitem [{\citenamefont {Mamin}\ \emph {et~al.}(2013)\citenamefont {Mamin},
  \citenamefont {Kim}, \citenamefont {Sherwood}, \citenamefont {Rettner},
  \citenamefont {Ohno}, \citenamefont {Awschalom},\ and\ \citenamefont
  {Rugar}}]{MKS+13}%
  \BibitemOpen
  \bibfield  {author} {\bibinfo {author} {\bibfnamefont {H.~J.}\ \bibnamefont
  {Mamin}}, \bibinfo {author} {\bibfnamefont {M.}~\bibnamefont {Kim}}, \bibinfo
  {author} {\bibfnamefont {M.~H.}\ \bibnamefont {Sherwood}}, \bibinfo {author}
  {\bibfnamefont {C.~T.}\ \bibnamefont {Rettner}}, \bibinfo {author}
  {\bibfnamefont {K.}~\bibnamefont {Ohno}}, \bibinfo {author} {\bibfnamefont
  {D.~D.}\ \bibnamefont {Awschalom}}, \ and\ \bibinfo {author} {\bibfnamefont
  {D.}~\bibnamefont {Rugar}},\ }\bibfield  {title} {\enquote {\bibinfo {title}
  {Nanoscale {N}uclear {M}agnetic {R}esonance with a {N}itrogen-{V}acancy
  {S}pin {S}ensor},}\ }\href@noop {} {\bibfield  {journal} {\bibinfo  {journal}
  {Science}\ }\textbf {\bibinfo {volume} {339}},\ \bibinfo {pages} {557}
  (\bibinfo {year} {2013})}\BibitemShut {NoStop}%
\bibitem [{\citenamefont {Staudacher}\ \emph {et~al.}(2013)\citenamefont
  {Staudacher}, \citenamefont {Shi}, \citenamefont {Pezzagna}, \citenamefont
  {Meijer}, \citenamefont {Du}, \citenamefont {Meriles}, \citenamefont
  {Reinhard},\ and\ \citenamefont {Wrachtrup}}]{SSP+13}%
  \BibitemOpen
  \bibfield  {author} {\bibinfo {author} {\bibfnamefont {T.}~\bibnamefont
  {Staudacher}}, \bibinfo {author} {\bibfnamefont {F.}~\bibnamefont {Shi}},
  \bibinfo {author} {\bibfnamefont {S.}~\bibnamefont {Pezzagna}}, \bibinfo
  {author} {\bibfnamefont {J.}~\bibnamefont {Meijer}}, \bibinfo {author}
  {\bibfnamefont {J.}~\bibnamefont {Du}}, \bibinfo {author} {\bibfnamefont
  {C.~A.}\ \bibnamefont {Meriles}}, \bibinfo {author} {\bibfnamefont
  {F.}~\bibnamefont {Reinhard}}, \ and\ \bibinfo {author} {\bibfnamefont
  {J.}~\bibnamefont {Wrachtrup}},\ }\bibfield  {title} {\enquote {\bibinfo
  {title} {Nuclear {M}agnetic {R}esonance {S}pectroscopy on a
  (5-{N}anometer)$^3$ {S}ample {V}olume},}\ }\href@noop {} {\bibfield
  {journal} {\bibinfo  {journal} {Science}\ }\textbf {\bibinfo {volume}
  {339}},\ \bibinfo {pages} {561} (\bibinfo {year} {2013})}\BibitemShut
  {NoStop}%
\bibitem [{\citenamefont {M{\"u}ller}\ \emph {et~al.}(2014)\citenamefont
  {M{\"u}ller}, \citenamefont {Kong}, \citenamefont {Cai}, \citenamefont
  {Melentijevi{\'c}}, \citenamefont {Stacey}, \citenamefont {Markham},
  \citenamefont {Twitchen}, \citenamefont {Isoya}, \citenamefont {Pezzagna},
  \citenamefont {Meijer}, \citenamefont {Du}, \citenamefont {Plenio},
  \citenamefont {Naydenov}, \citenamefont {McGuinness},\ and\ \citenamefont
  {Jelezko}}]{MKC+14}%
  \BibitemOpen
  \bibfield  {author} {\bibinfo {author} {\bibfnamefont {C.}~\bibnamefont
  {M{\"u}ller}}, \bibinfo {author} {\bibfnamefont {X.}~\bibnamefont {Kong}},
  \bibinfo {author} {\bibfnamefont {J.-M.}\ \bibnamefont {Cai}}, \bibinfo
  {author} {\bibfnamefont {K.}~\bibnamefont {Melentijevi{\'c}}}, \bibinfo
  {author} {\bibfnamefont {A.}~\bibnamefont {Stacey}}, \bibinfo {author}
  {\bibfnamefont {M.}~\bibnamefont {Markham}}, \bibinfo {author} {\bibfnamefont
  {D.}~\bibnamefont {Twitchen}}, \bibinfo {author} {\bibfnamefont
  {J.}~\bibnamefont {Isoya}}, \bibinfo {author} {\bibfnamefont
  {S.}~\bibnamefont {Pezzagna}}, \bibinfo {author} {\bibfnamefont
  {J.}~\bibnamefont {Meijer}}, \bibinfo {author} {\bibfnamefont {J.~F.}\
  \bibnamefont {Du}}, \bibinfo {author} {\bibfnamefont {M.~B.}\ \bibnamefont
  {Plenio}}, \bibinfo {author} {\bibfnamefont {B.}~\bibnamefont {Naydenov}},
  \bibinfo {author} {\bibfnamefont {L.~P.}\ \bibnamefont {McGuinness}}, \ and\
  \bibinfo {author} {\bibfnamefont {F.}~\bibnamefont {Jelezko}},\ }\bibfield
  {title} {\enquote {\bibinfo {title} {Nuclear magnetic resonance spectroscopy
  with single spin sensitivity},}\ }\href@noop {} {\bibfield  {journal}
  {\bibinfo  {journal} {Nat.\ Commun.}\ }\textbf {\bibinfo {volume} {5}},\
  \bibinfo {pages} {4703} (\bibinfo {year} {2014})}\BibitemShut {NoStop}%
\bibitem [{\citenamefont {H{\"a}berle}\ \emph {et~al.}(2015)\citenamefont
  {H{\"a}berle}, \citenamefont {Schmid-Lorch}, \citenamefont {Reinhard},\ and\
  \citenamefont {Wrachtrup}}]{HSR+15}%
  \BibitemOpen
  \bibfield  {author} {\bibinfo {author} {\bibfnamefont {T.}~\bibnamefont
  {H{\"a}berle}}, \bibinfo {author} {\bibfnamefont {D.}~\bibnamefont
  {Schmid-Lorch}}, \bibinfo {author} {\bibfnamefont {F.}~\bibnamefont
  {Reinhard}}, \ and\ \bibinfo {author} {\bibfnamefont {J.}~\bibnamefont
  {Wrachtrup}},\ }\bibfield  {title} {\enquote {\bibinfo {title} {Nanoscale
  nuclear magnetic imaging with chemical contrast},}\ }\href@noop {} {\bibfield
   {journal} {\bibinfo  {journal} {Nat.\ Nanotechnol.}\ }\textbf {\bibinfo
  {volume} {10}},\ \bibinfo {pages} {125} (\bibinfo {year} {2015})}\BibitemShut
  {NoStop}%
\bibitem [{\citenamefont {DeVience}\ \emph {et~al.}(2015)\citenamefont
  {DeVience}, \citenamefont {Pham}, \citenamefont {Lovchinsky}, \citenamefont
  {Sushkov}, \citenamefont {Bar-Gill}, \citenamefont {Belthangady},
  \citenamefont {Casola}, \citenamefont {Corbett}, \citenamefont {Zhang},
  \citenamefont {Lukin}, \citenamefont {Park}, \citenamefont {Yacoby},\ and\
  \citenamefont {Walsworth}}]{DPL+15}%
  \BibitemOpen
  \bibfield  {author} {\bibinfo {author} {\bibfnamefont {S.~J.}\ \bibnamefont
  {DeVience}}, \bibinfo {author} {\bibfnamefont {L.~M.}\ \bibnamefont {Pham}},
  \bibinfo {author} {\bibfnamefont {I.}~\bibnamefont {Lovchinsky}}, \bibinfo
  {author} {\bibfnamefont {A.~O.}\ \bibnamefont {Sushkov}}, \bibinfo {author}
  {\bibfnamefont {N.}~\bibnamefont {Bar-Gill}}, \bibinfo {author}
  {\bibfnamefont {C.}~\bibnamefont {Belthangady}}, \bibinfo {author}
  {\bibfnamefont {F.}~\bibnamefont {Casola}}, \bibinfo {author} {\bibfnamefont
  {M.}~\bibnamefont {Corbett}}, \bibinfo {author} {\bibfnamefont
  {H.}~\bibnamefont {Zhang}}, \bibinfo {author} {\bibfnamefont
  {M.}~\bibnamefont {Lukin}}, \bibinfo {author} {\bibfnamefont
  {H.}~\bibnamefont {Park}}, \bibinfo {author} {\bibfnamefont {A.}~\bibnamefont
  {Yacoby}}, \ and\ \bibinfo {author} {\bibfnamefont {R.~L.}\ \bibnamefont
  {Walsworth}},\ }\bibfield  {title} {\enquote {\bibinfo {title} {Nanoscale
  {NMR} spectroscopy and imaging of multiple nulcear species},}\ }\href@noop {}
  {\bibfield  {journal} {\bibinfo  {journal} {Nat.\ Nanotechnol.}\ }\textbf
  {\bibinfo {volume} {10}},\ \bibinfo {pages} {129} (\bibinfo {year}
  {2015})}\BibitemShut {NoStop}%
\bibitem [{\citenamefont {Aslam}\ \emph {et~al.}(2017)\citenamefont {Aslam},
  \citenamefont {Pfender}, \citenamefont {Neumann}, \citenamefont {Reuter},
  \citenamefont {Zappe}, \citenamefont {de~Oliveira}, \citenamefont
  {Denisenko}, \citenamefont {Sumiya}, \citenamefont {Onoda}, \citenamefont
  {Isoya},\ and\ \citenamefont {Wrachtrup}}]{APN+17}%
  \BibitemOpen
  \bibfield  {author} {\bibinfo {author} {\bibfnamefont {N.}~\bibnamefont
  {Aslam}}, \bibinfo {author} {\bibfnamefont {M.}~\bibnamefont {Pfender}},
  \bibinfo {author} {\bibfnamefont {P.}~\bibnamefont {Neumann}}, \bibinfo
  {author} {\bibfnamefont {R.}~\bibnamefont {Reuter}}, \bibinfo {author}
  {\bibfnamefont {A.}~\bibnamefont {Zappe}}, \bibinfo {author} {\bibfnamefont
  {F.~F.}\ \bibnamefont {de~Oliveira}}, \bibinfo {author} {\bibfnamefont
  {A.}~\bibnamefont {Denisenko}}, \bibinfo {author} {\bibfnamefont
  {H.}~\bibnamefont {Sumiya}}, \bibinfo {author} {\bibfnamefont
  {S.}~\bibnamefont {Onoda}}, \bibinfo {author} {\bibfnamefont
  {J.}~\bibnamefont {Isoya}}, \ and\ \bibinfo {author} {\bibfnamefont
  {J.}~\bibnamefont {Wrachtrup}},\ }\bibfield  {title} {\enquote {\bibinfo
  {title} {Nanoscale nuclear magnetic resonance with chemical resolution},}\
  }\href@noop {} {\bibfield  {journal} {\bibinfo  {journal} {Science}\ }\textbf
  {\bibinfo {volume} {357}},\ \bibinfo {pages} {67} (\bibinfo {year}
  {2017})}\BibitemShut {NoStop}%
\bibitem [{\citenamefont {Lovchinsky}\ \emph {et~al.}(2016)\citenamefont
  {Lovchinsky}, \citenamefont {Sushkov}, \citenamefont {Urbach}, \citenamefont
  {de~Leon}, \citenamefont {Choi}, \citenamefont {De~Greve}, \citenamefont
  {Evans}, \citenamefont {Gertner}, \citenamefont {Bersin}, \citenamefont
  {M{\"u}ller}, \citenamefont {McGuinness}, \citenamefont {Jelezko},
  \citenamefont {Walsworth}, \citenamefont {Park},\ and\ \citenamefont
  {Lukin}}]{LSU+16}%
  \BibitemOpen
  \bibfield  {author} {\bibinfo {author} {\bibfnamefont {I.}~\bibnamefont
  {Lovchinsky}}, \bibinfo {author} {\bibfnamefont {A.~O.}\ \bibnamefont
  {Sushkov}}, \bibinfo {author} {\bibfnamefont {E.}~\bibnamefont {Urbach}},
  \bibinfo {author} {\bibfnamefont {N.~P.}\ \bibnamefont {de~Leon}}, \bibinfo
  {author} {\bibfnamefont {S.}~\bibnamefont {Choi}}, \bibinfo {author}
  {\bibfnamefont {K.}~\bibnamefont {De~Greve}}, \bibinfo {author}
  {\bibfnamefont {R.}~\bibnamefont {Evans}}, \bibinfo {author} {\bibfnamefont
  {R.}~\bibnamefont {Gertner}}, \bibinfo {author} {\bibfnamefont
  {E.}~\bibnamefont {Bersin}}, \bibinfo {author} {\bibfnamefont
  {C.}~\bibnamefont {M{\"u}ller}}, \bibinfo {author} {\bibfnamefont
  {L.}~\bibnamefont {McGuinness}}, \bibinfo {author} {\bibfnamefont
  {F.}~\bibnamefont {Jelezko}}, \bibinfo {author} {\bibfnamefont {R.~L.}\
  \bibnamefont {Walsworth}}, \bibinfo {author} {\bibfnamefont {H.}~\bibnamefont
  {Park}}, \ and\ \bibinfo {author} {\bibfnamefont {M.~D.}\ \bibnamefont
  {Lukin}},\ }\bibfield  {title} {\enquote {\bibinfo {title} {Nuclear magnetic
  resonance detection and spectroscopy of single proteins using quantum
  logic},}\ }\href@noop {} {\bibfield  {journal} {\bibinfo  {journal}
  {Science}\ }\textbf {\bibinfo {volume} {351}},\ \bibinfo {pages} {836}
  (\bibinfo {year} {2016})}\BibitemShut {NoStop}%
\bibitem [{\citenamefont {Sushkov}\ \emph {et~al.}(2014)\citenamefont
  {Sushkov}, \citenamefont {Lovchinsky}, \citenamefont {Chisholm},
  \citenamefont {Walsworth}, \citenamefont {Park},\ and\ \citenamefont
  {Lukin}}]{SLC+14}%
  \BibitemOpen
  \bibfield  {author} {\bibinfo {author} {\bibfnamefont {A.~O.}\ \bibnamefont
  {Sushkov}}, \bibinfo {author} {\bibfnamefont {I.}~\bibnamefont {Lovchinsky}},
  \bibinfo {author} {\bibfnamefont {N.}~\bibnamefont {Chisholm}}, \bibinfo
  {author} {\bibfnamefont {R.~L.}\ \bibnamefont {Walsworth}}, \bibinfo {author}
  {\bibfnamefont {H.}~\bibnamefont {Park}}, \ and\ \bibinfo {author}
  {\bibfnamefont {M.~D.}\ \bibnamefont {Lukin}},\ }\bibfield  {title} {\enquote
  {\bibinfo {title} {Magnetic {R}esonance {D}etection of {I}ndividual {P}roton
  {S}pins {U}sing {Q}uantum {R}eporters},}\ }\href@noop {} {\bibfield
  {journal} {\bibinfo  {journal} {Phys.\ Rev.\ Lett.}\ }\textbf {\bibinfo
  {volume} {113}},\ \bibinfo {pages} {197601} (\bibinfo {year}
  {2014})}\BibitemShut {NoStop}%
\bibitem [{\citenamefont {Maurer}\ \emph {et~al.}(2012)\citenamefont {Maurer},
  \citenamefont {Kucsko}, \citenamefont {Latta}, \citenamefont {Jiang},
  \citenamefont {Yao}, \citenamefont {Bennett}, \citenamefont {Pastawski},
  \citenamefont {Hunger}, \citenamefont {Chisholm}, \citenamefont {Markham},
  \citenamefont {Twitchen}, \citenamefont {Cirac},\ and\ \citenamefont
  {Lukin}}]{MKL+12}%
  \BibitemOpen
  \bibfield  {author} {\bibinfo {author} {\bibfnamefont {P.~C.}\ \bibnamefont
  {Maurer}}, \bibinfo {author} {\bibfnamefont {G.}~\bibnamefont {Kucsko}},
  \bibinfo {author} {\bibfnamefont {C.}~\bibnamefont {Latta}}, \bibinfo
  {author} {\bibfnamefont {L.}~\bibnamefont {Jiang}}, \bibinfo {author}
  {\bibfnamefont {N.~Y.}\ \bibnamefont {Yao}}, \bibinfo {author} {\bibfnamefont
  {S.~D.}\ \bibnamefont {Bennett}}, \bibinfo {author} {\bibfnamefont
  {F.}~\bibnamefont {Pastawski}}, \bibinfo {author} {\bibfnamefont
  {D.}~\bibnamefont {Hunger}}, \bibinfo {author} {\bibfnamefont
  {N.}~\bibnamefont {Chisholm}}, \bibinfo {author} {\bibfnamefont
  {M.}~\bibnamefont {Markham}}, \bibinfo {author} {\bibfnamefont {D.~J.}\
  \bibnamefont {Twitchen}}, \bibinfo {author} {\bibfnamefont {J.~I.}\
  \bibnamefont {Cirac}}, \ and\ \bibinfo {author} {\bibfnamefont {M.~D.}\
  \bibnamefont {Lukin}},\ }\bibfield  {title} {\enquote {\bibinfo {title}
  {Maurer room-{T}emperature {Q}uantum {B}it {M}emory {E}xceeding {O}ne
  {S}econd},}\ }\href@noop {} {\bibfield  {journal} {\bibinfo  {journal}
  {Science}\ }\textbf {\bibinfo {volume} {336}},\ \bibinfo {pages} {1283}
  (\bibinfo {year} {2012})}\BibitemShut {NoStop}%
\bibitem [{\citenamefont {Bradley}\ \emph {et~al.}(2019)\citenamefont
  {Bradley}, \citenamefont {Randall}, \citenamefont {Abobeih}, \citenamefont
  {Berrevoets}, \citenamefont {Degen}, \citenamefont {Bakker}, \citenamefont
  {Markham}, \citenamefont {Twitchen},\ and\ \citenamefont
  {Taminiau}}]{BRA+19}%
  \BibitemOpen
  \bibfield  {author} {\bibinfo {author} {\bibfnamefont {C.~E.}\ \bibnamefont
  {Bradley}}, \bibinfo {author} {\bibfnamefont {J.}~\bibnamefont {Randall}},
  \bibinfo {author} {\bibfnamefont {M.~H.}\ \bibnamefont {Abobeih}}, \bibinfo
  {author} {\bibfnamefont {R.~C.}\ \bibnamefont {Berrevoets}}, \bibinfo
  {author} {\bibfnamefont {M.~J.}\ \bibnamefont {Degen}}, \bibinfo {author}
  {\bibfnamefont {M.~A.}\ \bibnamefont {Bakker}}, \bibinfo {author}
  {\bibfnamefont {M.}~\bibnamefont {Markham}}, \bibinfo {author} {\bibfnamefont
  {D.~J.}\ \bibnamefont {Twitchen}}, \ and\ \bibinfo {author} {\bibfnamefont
  {T.~H.}\ \bibnamefont {Taminiau}},\ }\bibfield  {title} {\enquote {\bibinfo
  {title} {A {T}en-{Q}ubit {S}olid-{S}tate {S}pin {R}egister with {Q}uantum
  {M}emory up to {O}ne {M}inute},}\ }\href@noop {} {\bibfield  {journal}
  {\bibinfo  {journal} {Phys.\ Rev.\ X}\ }\textbf {\bibinfo {volume} {9}},\
  \bibinfo {pages} {031045} (\bibinfo {year} {2019})}\BibitemShut {NoStop}%
\bibitem [{\citenamefont {Sumiya}\ and\ \citenamefont {Satoh}(1996)}]{SS96}%
  \BibitemOpen
  \bibfield  {author} {\bibinfo {author} {\bibfnamefont {H.}~\bibnamefont
  {Sumiya}}\ and\ \bibinfo {author} {\bibfnamefont {S.}~\bibnamefont {Satoh}},\
  }\bibfield  {title} {\enquote {\bibinfo {title} {High-pressure synthesis of
  high-purity diamond crystal},}\ }\href@noop {} {\bibfield  {journal}
  {\bibinfo  {journal} {Diam.\ Relat.\ Mater.}\ }\textbf {\bibinfo {volume}
  {5}},\ \bibinfo {pages} {1359} (\bibinfo {year} {1996})}\BibitemShut
  {NoStop}%
\bibitem [{\citenamefont {Watanabe}\ \emph {et~al.}(1999)\citenamefont
  {Watanabe}, \citenamefont {Takeuchi}, \citenamefont {Yamanaka}, \citenamefont
  {Okushi}, \citenamefont {Kajimura},\ and\ \citenamefont {Sekiguch}}]{WTY+99}%
  \BibitemOpen
  \bibfield  {author} {\bibinfo {author} {\bibfnamefont {H.}~\bibnamefont
  {Watanabe}}, \bibinfo {author} {\bibfnamefont {D.}~\bibnamefont {Takeuchi}},
  \bibinfo {author} {\bibfnamefont {S.}~\bibnamefont {Yamanaka}}, \bibinfo
  {author} {\bibfnamefont {H.}~\bibnamefont {Okushi}}, \bibinfo {author}
  {\bibfnamefont {K.}~\bibnamefont {Kajimura}}, \ and\ \bibinfo {author}
  {\bibfnamefont {T.}~\bibnamefont {Sekiguch}},\ }\bibfield  {title} {\enquote
  {\bibinfo {title} {Homoepitaxial diamond film with an atomically flat surface
  over a large area},}\ }\href@noop {} {\bibfield  {journal} {\bibinfo
  {journal} {Diam.\ Relat.\ Mater.}\ }\textbf {\bibinfo {volume} {8}},\
  \bibinfo {pages} {1272} (\bibinfo {year} {1999})}\BibitemShut {NoStop}%
\bibitem [{\citenamefont {Itoh}\ and\ \citenamefont {Watanabe}(2014)}]{IW14}%
  \BibitemOpen
  \bibfield  {author} {\bibinfo {author} {\bibfnamefont {K.~M.}\ \bibnamefont
  {Itoh}}\ and\ \bibinfo {author} {\bibfnamefont {H.}~\bibnamefont
  {Watanabe}},\ }\bibfield  {title} {\enquote {\bibinfo {title} {Isotope
  engineering of silicon and diamond for quantum computing and sensing
  applications},}\ }\href@noop {} {\bibfield  {journal} {\bibinfo  {journal}
  {MRS Commun.}\ }\textbf {\bibinfo {volume} {4}},\ \bibinfo {pages} {143}
  (\bibinfo {year} {2014})}\BibitemShut {NoStop}%
\bibitem [{\citenamefont {Abe}\ and\ \citenamefont {Sasaki}(2018)}]{AS18}%
  \BibitemOpen
  \bibfield  {author} {\bibinfo {author} {\bibfnamefont {E.}~\bibnamefont
  {Abe}}\ and\ \bibinfo {author} {\bibfnamefont {K.}~\bibnamefont {Sasaki}},\
  }\bibfield  {title} {\enquote {\bibinfo {title} {Tutorial: {M}agnetic
  resonance with nitrogen-vacancy centers in diamond---microwave engineering,
  materials science, and magnetometry},}\ }\href@noop {} {\bibfield  {journal}
  {\bibinfo  {journal} {J.\ Appl.\ Phys.}\ }\textbf {\bibinfo {volume} {123}},\
  \bibinfo {pages} {161101} (\bibinfo {year} {2018})}\BibitemShut {NoStop}%
\bibitem [{\citenamefont {Sasaki}, \citenamefont {Itoh},\ and\ \citenamefont
  {Abe}(2018)}]{SIA18}%
  \BibitemOpen
  \bibfield  {author} {\bibinfo {author} {\bibfnamefont {K.}~\bibnamefont
  {Sasaki}}, \bibinfo {author} {\bibfnamefont {K.~M.}\ \bibnamefont {Itoh}}, \
  and\ \bibinfo {author} {\bibfnamefont {E.}~\bibnamefont {Abe}},\ }\bibfield
  {title} {\enquote {\bibinfo {title} {Determination of the position of a
  single nuclear spin from free nuclear precessions detected by a solid-state
  quantum sensor},}\ }\href@noop {} {\bibfield  {journal} {\bibinfo  {journal}
  {Phys.\ Rev.\ B}\ }\textbf {\bibinfo {volume} {98}},\ \bibinfo {pages}
  {121405} (\bibinfo {year} {2018})}\BibitemShut {NoStop}%
\bibitem [{\citenamefont {Gullion}, \citenamefont {Baker},\ and\ \citenamefont
  {Conradi}(1990)}]{GBC90}%
  \BibitemOpen
  \bibfield  {author} {\bibinfo {author} {\bibfnamefont {T.}~\bibnamefont
  {Gullion}}, \bibinfo {author} {\bibfnamefont {D.~B.}\ \bibnamefont {Baker}},
  \ and\ \bibinfo {author} {\bibfnamefont {M.~S.}\ \bibnamefont {Conradi}},\
  }\bibfield  {title} {\enquote {\bibinfo {title} {New, compensated
  {C}arr-{P}urcell sequences},}\ }\href@noop {} {\bibfield  {journal} {\bibinfo
   {journal} {J.\ Mag.\ Res.}\ }\textbf {\bibinfo {volume} {89}},\ \bibinfo
  {pages} {479} (\bibinfo {year} {1990})}\BibitemShut {NoStop}%
\bibitem [{\citenamefont {Kolkowitz}\ \emph {et~al.}(2012)\citenamefont
  {Kolkowitz}, \citenamefont {Unterreithmeier}, \citenamefont {Bennett},\ and\
  \citenamefont {Lukin}}]{KUBL12}%
  \BibitemOpen
  \bibfield  {author} {\bibinfo {author} {\bibfnamefont {S.}~\bibnamefont
  {Kolkowitz}}, \bibinfo {author} {\bibfnamefont {Q.~P.}\ \bibnamefont
  {Unterreithmeier}}, \bibinfo {author} {\bibfnamefont {S.~D.}\ \bibnamefont
  {Bennett}}, \ and\ \bibinfo {author} {\bibfnamefont {M.~D.}\ \bibnamefont
  {Lukin}},\ }\bibfield  {title} {\enquote {\bibinfo {title} {Sensing {D}istant
  {N}uclear {S}pins with a {S}ingle {E}lectron {S}pin},}\ }\href@noop {}
  {\bibfield  {journal} {\bibinfo  {journal} {Phys.\ Rev.\ Lett.}\ }\textbf
  {\bibinfo {volume} {109}},\ \bibinfo {pages} {137601} (\bibinfo {year}
  {2012})}\BibitemShut {NoStop}%
\bibitem [{\citenamefont {Taminiau}\ \emph {et~al.}(2012)\citenamefont
  {Taminiau}, \citenamefont {Wagenaar}, \citenamefont {van~der Sar},
  \citenamefont {Jelezko}, \citenamefont {Dobrovitski},\ and\ \citenamefont
  {Hanson}}]{TWS+12}%
  \BibitemOpen
  \bibfield  {author} {\bibinfo {author} {\bibfnamefont {T.~H.}\ \bibnamefont
  {Taminiau}}, \bibinfo {author} {\bibfnamefont {J.~J.~T.}\ \bibnamefont
  {Wagenaar}}, \bibinfo {author} {\bibfnamefont {T.}~\bibnamefont {van~der
  Sar}}, \bibinfo {author} {\bibfnamefont {F.}~\bibnamefont {Jelezko}},
  \bibinfo {author} {\bibfnamefont {V.~V.}\ \bibnamefont {Dobrovitski}}, \ and\
  \bibinfo {author} {\bibfnamefont {R.}~\bibnamefont {Hanson}},\ }\bibfield
  {title} {\enquote {\bibinfo {title} {Detection and {C}ontrol of {I}ndividual
  {N}uclear {S}pin {U}sing a {W}eakly {C}oupled {E}lectron {S}pin},}\
  }\href@noop {} {\bibfield  {journal} {\bibinfo  {journal} {Phys.\ Rev.\
  Lett.}\ }\textbf {\bibinfo {volume} {109}},\ \bibinfo {pages} {137602}
  (\bibinfo {year} {2012})}\BibitemShut {NoStop}%
\bibitem [{\citenamefont {Boss}\ \emph {et~al.}(2016)\citenamefont {Boss},
  \citenamefont {Chang}, \citenamefont {Armijo}, \citenamefont {Cujia},
  \citenamefont {Rosskopf}, \citenamefont {Maze},\ and\ \citenamefont
  {Degen}}]{BCA+16}%
  \BibitemOpen
  \bibfield  {author} {\bibinfo {author} {\bibfnamefont {J.~M.}\ \bibnamefont
  {Boss}}, \bibinfo {author} {\bibfnamefont {K.}~\bibnamefont {Chang}},
  \bibinfo {author} {\bibfnamefont {J.}~\bibnamefont {Armijo}}, \bibinfo
  {author} {\bibfnamefont {K.}~\bibnamefont {Cujia}}, \bibinfo {author}
  {\bibfnamefont {T.}~\bibnamefont {Rosskopf}}, \bibinfo {author}
  {\bibfnamefont {J.~R.}\ \bibnamefont {Maze}}, \ and\ \bibinfo {author}
  {\bibfnamefont {C.~L.}\ \bibnamefont {Degen}},\ }\bibfield  {title} {\enquote
  {\bibinfo {title} {One- and {T}wo-{D}imensional {N}uclear {M}agnetic
  {R}esonance {S}pectroscopy with a {D}iamond {Q}uantum {S}ensor},}\
  }\href@noop {} {\bibfield  {journal} {\bibinfo  {journal} {Phys.\ Rev.\
  Lett.}\ }\textbf {\bibinfo {volume} {116}},\ \bibinfo {pages} {197601}
  (\bibinfo {year} {2016})}\BibitemShut {NoStop}%
\bibitem [{\citenamefont {Misonou}\ \emph {et~al.}(2020)\citenamefont
  {Misonou}, \citenamefont {Sasaki}, \citenamefont {Ishizu}, \citenamefont
  {Monnai}, \citenamefont {Itoh},\ and\ \citenamefont {Abe}}]{MSI+20}%
  \BibitemOpen
  \bibfield  {author} {\bibinfo {author} {\bibfnamefont {D.}~\bibnamefont
  {Misonou}}, \bibinfo {author} {\bibfnamefont {K.}~\bibnamefont {Sasaki}},
  \bibinfo {author} {\bibfnamefont {S.}~\bibnamefont {Ishizu}}, \bibinfo
  {author} {\bibfnamefont {Y.}~\bibnamefont {Monnai}}, \bibinfo {author}
  {\bibfnamefont {K.~M.}\ \bibnamefont {Itoh}}, \ and\ \bibinfo {author}
  {\bibfnamefont {E.}~\bibnamefont {Abe}},\ }\bibfield  {title} {\enquote
  {\bibinfo {title} {Construction and operation of a tabletop system for
  nanoscale magnetometry with single nitrogen-vacancy centers in diamond},}\
  }\href@noop {} {\bibfield  {journal} {\bibinfo  {journal} {AIP Adv.}\
  }\textbf {\bibinfo {volume} {10}},\ \bibinfo {pages} {025206} (\bibinfo
  {year} {2020})}\BibitemShut {NoStop}%
\bibitem [{\citenamefont {Zhao}\ \emph {et~al.}(2011)\citenamefont {Zhao},
  \citenamefont {Hu}, \citenamefont {Ho}, \citenamefont {Wan},\ and\
  \citenamefont {Liu}}]{ZHH+11}%
  \BibitemOpen
  \bibfield  {author} {\bibinfo {author} {\bibfnamefont {N.}~\bibnamefont
  {Zhao}}, \bibinfo {author} {\bibfnamefont {J.-L.}\ \bibnamefont {Hu}},
  \bibinfo {author} {\bibfnamefont {S.-W.}\ \bibnamefont {Ho}}, \bibinfo
  {author} {\bibfnamefont {J.~T.~K.}\ \bibnamefont {Wan}}, \ and\ \bibinfo
  {author} {\bibfnamefont {R.~B.}\ \bibnamefont {Liu}},\ }\bibfield  {title}
  {\enquote {\bibinfo {title} {Atomic-scale magnetometry of distant nuclear
  spin cluster via nitrogen-vacancy spin in diamond},}\ }\href@noop {}
  {\bibfield  {journal} {\bibinfo  {journal} {Nat.\ Nanotechnol.}\ }\textbf
  {\bibinfo {volume} {6}},\ \bibinfo {pages} {242} (\bibinfo {year}
  {2011})}\BibitemShut {NoStop}%
\bibitem [{\citenamefont {Zhao}\ \emph {et~al.}(2012)\citenamefont {Zhao},
  \citenamefont {Honert}, \citenamefont {Schmid}, \citenamefont {Klas},
  \citenamefont {Isoya}, \citenamefont {Markham}, \citenamefont {Twitchen},
  \citenamefont {Jelezko}, \citenamefont {Liu}, \citenamefont {Fedder},\ and\
  \citenamefont {Wrachtrup}}]{ZHS+12}%
  \BibitemOpen
  \bibfield  {author} {\bibinfo {author} {\bibfnamefont {N.}~\bibnamefont
  {Zhao}}, \bibinfo {author} {\bibfnamefont {J.}~\bibnamefont {Honert}},
  \bibinfo {author} {\bibfnamefont {B.}~\bibnamefont {Schmid}}, \bibinfo
  {author} {\bibfnamefont {M.}~\bibnamefont {Klas}}, \bibinfo {author}
  {\bibfnamefont {J.}~\bibnamefont {Isoya}}, \bibinfo {author} {\bibfnamefont
  {M.}~\bibnamefont {Markham}}, \bibinfo {author} {\bibfnamefont
  {D.}~\bibnamefont {Twitchen}}, \bibinfo {author} {\bibfnamefont
  {F.}~\bibnamefont {Jelezko}}, \bibinfo {author} {\bibfnamefont {R.~B.}\
  \bibnamefont {Liu}}, \bibinfo {author} {\bibfnamefont {H.}~\bibnamefont
  {Fedder}}, \ and\ \bibinfo {author} {\bibfnamefont {J.}~\bibnamefont
  {Wrachtrup}},\ }\bibfield  {title} {\enquote {\bibinfo {title} {Sensing
  single remote nuclear spins},}\ }\href@noop {} {\bibfield  {journal}
  {\bibinfo  {journal} {Nat.\ Nanotechnol.}\ }\textbf {\bibinfo {volume} {7}},\
  \bibinfo {pages} {657} (\bibinfo {year} {2012})}\BibitemShut {NoStop}%
\bibitem [{\citenamefont {Laraoui}\ \emph {et~al.}(2013)\citenamefont
  {Laraoui}, \citenamefont {Dolde}, \citenamefont {Burk}, \citenamefont
  {Reinhard}, \citenamefont {Wrachtrup},\ and\ \citenamefont
  {Meriles}}]{LDB+13}%
  \BibitemOpen
  \bibfield  {author} {\bibinfo {author} {\bibfnamefont {A.}~\bibnamefont
  {Laraoui}}, \bibinfo {author} {\bibfnamefont {F.}~\bibnamefont {Dolde}},
  \bibinfo {author} {\bibfnamefont {C.}~\bibnamefont {Burk}}, \bibinfo {author}
  {\bibfnamefont {F.}~\bibnamefont {Reinhard}}, \bibinfo {author}
  {\bibfnamefont {J.}~\bibnamefont {Wrachtrup}}, \ and\ \bibinfo {author}
  {\bibfnamefont {C.~A.}\ \bibnamefont {Meriles}},\ }\bibfield  {title}
  {\enquote {\bibinfo {title} {High-resolution correlation spectroscopy of
  $^{13}${C} spins near a nitrogen vacancy centre in diamond},}\ }\href@noop {}
  {\bibfield  {journal} {\bibinfo  {journal} {Nat.\ Commun.}\ }\textbf
  {\bibinfo {volume} {4}},\ \bibinfo {pages} {1651} (\bibinfo {year}
  {2013})}\BibitemShut {NoStop}%
\bibitem [{\citenamefont {Kong}\ \emph {et~al.}(2015)\citenamefont {Kong},
  \citenamefont {Stark}, \citenamefont {Du}, \citenamefont {McGuinness},\ and\
  \citenamefont {Jelezko}}]{KSD+15}%
  \BibitemOpen
  \bibfield  {author} {\bibinfo {author} {\bibfnamefont {X.}~\bibnamefont
  {Kong}}, \bibinfo {author} {\bibfnamefont {A.}~\bibnamefont {Stark}},
  \bibinfo {author} {\bibfnamefont {J.}~\bibnamefont {Du}}, \bibinfo {author}
  {\bibfnamefont {L.~P.}\ \bibnamefont {McGuinness}}, \ and\ \bibinfo {author}
  {\bibfnamefont {F.}~\bibnamefont {Jelezko}},\ }\bibfield  {title} {\enquote
  {\bibinfo {title} {Towards {C}hemical {S}tructure {R}esolution with
  {N}anoscale {N}uclear {M}agnetic {R}esonance {S}pectroscopy},}\ }\href@noop
  {} {\bibfield  {journal} {\bibinfo  {journal} {Phys.\ Rev.\ Appl.}\ }\textbf
  {\bibinfo {volume} {4}},\ \bibinfo {pages} {024004} (\bibinfo {year}
  {2015})}\BibitemShut {NoStop}%
\bibitem [{\citenamefont {Staudacher}\ \emph {et~al.}(2015)\citenamefont
  {Staudacher}, \citenamefont {Raatz}, \citenamefont {Pezzagna}, \citenamefont
  {Meijer}, \citenamefont {Reinhard}, \citenamefont {Meriles},\ and\
  \citenamefont {Wrachtrup}}]{SRP+15}%
  \BibitemOpen
  \bibfield  {author} {\bibinfo {author} {\bibfnamefont {T.}~\bibnamefont
  {Staudacher}}, \bibinfo {author} {\bibfnamefont {N.}~\bibnamefont {Raatz}},
  \bibinfo {author} {\bibfnamefont {S.}~\bibnamefont {Pezzagna}}, \bibinfo
  {author} {\bibfnamefont {J.}~\bibnamefont {Meijer}}, \bibinfo {author}
  {\bibfnamefont {F.}~\bibnamefont {Reinhard}}, \bibinfo {author}
  {\bibfnamefont {C.~A.}\ \bibnamefont {Meriles}}, \ and\ \bibinfo {author}
  {\bibfnamefont {J.}~\bibnamefont {Wrachtrup}},\ }\bibfield  {title} {\enquote
  {\bibinfo {title} {Probing molecular dynamics at the nanoscale via an
  individual paramagnetic centre},}\ }\href@noop {} {\bibfield  {journal}
  {\bibinfo  {journal} {Nat.\ Commun.}\ }\textbf {\bibinfo {volume} {6}},\
  \bibinfo {pages} {8527} (\bibinfo {year} {2015})}\BibitemShut {NoStop}%
\bibitem [{\citenamefont {Loretz}\ \emph {et~al.}(2015)\citenamefont {Loretz},
  \citenamefont {Boss}, \citenamefont {Rosskopf}, \citenamefont {Mamin},
  \citenamefont {Rugar},\ and\ \citenamefont {Degen}}]{LBR+15}%
  \BibitemOpen
  \bibfield  {author} {\bibinfo {author} {\bibfnamefont {M.}~\bibnamefont
  {Loretz}}, \bibinfo {author} {\bibfnamefont {J.~M.}\ \bibnamefont {Boss}},
  \bibinfo {author} {\bibfnamefont {T.}~\bibnamefont {Rosskopf}}, \bibinfo
  {author} {\bibfnamefont {H.~J.}\ \bibnamefont {Mamin}}, \bibinfo {author}
  {\bibfnamefont {D.}~\bibnamefont {Rugar}}, \ and\ \bibinfo {author}
  {\bibfnamefont {C.~L.}\ \bibnamefont {Degen}},\ }\bibfield  {title} {\enquote
  {\bibinfo {title} {Spurious {H}armonic {R}esponse of {M}ultipulse {Q}uantum
  {S}ensing {S}equences},}\ }\href@noop {} {\bibfield  {journal} {\bibinfo
  {journal} {Phys.\ Rev.\ X}\ }\textbf {\bibinfo {volume} {5}},\ \bibinfo
  {pages} {021009} (\bibinfo {year} {2015})}\BibitemShut {NoStop}%
\bibitem [{\citenamefont {Zopes}\ \emph
  {et~al.}(2018{\natexlab{a}})\citenamefont {Zopes}, \citenamefont {Cujia},
  \citenamefont {Sasaki}, \citenamefont {Boss}, \citenamefont {Itoh},\ and\
  \citenamefont {Degen}}]{ZCS+18}%
  \BibitemOpen
  \bibfield  {author} {\bibinfo {author} {\bibfnamefont {J.}~\bibnamefont
  {Zopes}}, \bibinfo {author} {\bibfnamefont {K.~S.}\ \bibnamefont {Cujia}},
  \bibinfo {author} {\bibfnamefont {K.}~\bibnamefont {Sasaki}}, \bibinfo
  {author} {\bibfnamefont {J.~M.}\ \bibnamefont {Boss}}, \bibinfo {author}
  {\bibfnamefont {K.~M.}\ \bibnamefont {Itoh}}, \ and\ \bibinfo {author}
  {\bibfnamefont {C.~L.}\ \bibnamefont {Degen}},\ }\bibfield  {title} {\enquote
  {\bibinfo {title} {Three-dimensional localization spectroscopy of individual
  nuclear spins with sub-{A}ngstrom resolution},}\ }\href@noop {} {\bibfield
  {journal} {\bibinfo  {journal} {Nat.\ Commun.}\ }\textbf {\bibinfo {volume}
  {9}},\ \bibinfo {pages} {4678} (\bibinfo {year}
  {2018}{\natexlab{a}})}\BibitemShut {NoStop}%
\bibitem [{\citenamefont {Zopes}\ \emph
  {et~al.}(2018{\natexlab{b}})\citenamefont {Zopes}, \citenamefont {Herb},
  \citenamefont {Cujia},\ and\ \citenamefont {Degen}}]{ZHCD18}%
  \BibitemOpen
  \bibfield  {author} {\bibinfo {author} {\bibfnamefont {J.}~\bibnamefont
  {Zopes}}, \bibinfo {author} {\bibfnamefont {K.}~\bibnamefont {Herb}},
  \bibinfo {author} {\bibfnamefont {K.~S.}\ \bibnamefont {Cujia}}, \ and\
  \bibinfo {author} {\bibfnamefont {C.~L.}\ \bibnamefont {Degen}},\ }\bibfield
  {title} {\enquote {\bibinfo {title} {Three-{D}imensional {N}uclear {S}pin
  {P}ositioning {U}sing {C}oherent {R}adio-{F}requency {C}ontrol},}\
  }\href@noop {} {\bibfield  {journal} {\bibinfo  {journal} {Phys.\ Rev.\
  Lett.}\ }\textbf {\bibinfo {volume} {121}},\ \bibinfo {pages} {170801}
  (\bibinfo {year} {2018}{\natexlab{b}})}\BibitemShut {NoStop}%
\bibitem [{\citenamefont {Abobeih}\ \emph {et~al.}(2019)\citenamefont
  {Abobeih}, \citenamefont {Randall}, \citenamefont {Bradley}, \citenamefont
  {Bartling}, \citenamefont {Bakker}, \citenamefont {Degen}, \citenamefont
  {Markham}, \citenamefont {Twitchen},\ and\ \citenamefont
  {Taminiau}}]{ARB+19}%
  \BibitemOpen
  \bibfield  {author} {\bibinfo {author} {\bibfnamefont {M.~H.}\ \bibnamefont
  {Abobeih}}, \bibinfo {author} {\bibfnamefont {J.}~\bibnamefont {Randall}},
  \bibinfo {author} {\bibfnamefont {C.~E.}\ \bibnamefont {Bradley}}, \bibinfo
  {author} {\bibfnamefont {H.~P.}\ \bibnamefont {Bartling}}, \bibinfo {author}
  {\bibfnamefont {M.~A.}\ \bibnamefont {Bakker}}, \bibinfo {author}
  {\bibfnamefont {M.~J.}\ \bibnamefont {Degen}}, \bibinfo {author}
  {\bibfnamefont {M.}~\bibnamefont {Markham}}, \bibinfo {author} {\bibfnamefont
  {D.~J.}\ \bibnamefont {Twitchen}}, \ and\ \bibinfo {author} {\bibfnamefont
  {T.~H.}\ \bibnamefont {Taminiau}},\ }\bibfield  {title} {\enquote {\bibinfo
  {title} {Atomic-scale imaging of a 27-nuclear-spin cluster using a quantum
  sensor},}\ }\href@noop {} {\bibfield  {journal} {\bibinfo  {journal}
  {Nature}\ }\textbf {\bibinfo {volume} {576}},\ \bibinfo {pages} {411}
  (\bibinfo {year} {2019})}\BibitemShut {NoStop}%
\bibitem [{\citenamefont {Schwartz}\ \emph {et~al.}(2018)\citenamefont
  {Schwartz}, \citenamefont {Scheuer}, \citenamefont {Tratzmiller},
  \citenamefont {M{\"u}ller}, \citenamefont {Chen}, \citenamefont {Dhand},
  \citenamefont {Wang}, \citenamefont {M{\"u}ller}, \citenamefont {Naydenov},
  \citenamefont {Jelezko},\ and\ \citenamefont {Plenio}}]{SST+18}%
  \BibitemOpen
  \bibfield  {author} {\bibinfo {author} {\bibfnamefont {I.}~\bibnamefont
  {Schwartz}}, \bibinfo {author} {\bibfnamefont {J.}~\bibnamefont {Scheuer}},
  \bibinfo {author} {\bibfnamefont {B.}~\bibnamefont {Tratzmiller}}, \bibinfo
  {author} {\bibfnamefont {S.}~\bibnamefont {M{\"u}ller}}, \bibinfo {author}
  {\bibfnamefont {Q.}~\bibnamefont {Chen}}, \bibinfo {author} {\bibfnamefont
  {I.}~\bibnamefont {Dhand}}, \bibinfo {author} {\bibfnamefont
  {Z.}~\bibnamefont {Wang}}, \bibinfo {author} {\bibfnamefont {C.}~\bibnamefont
  {M{\"u}ller}}, \bibinfo {author} {\bibfnamefont {B.}~\bibnamefont
  {Naydenov}}, \bibinfo {author} {\bibfnamefont {F.}~\bibnamefont {Jelezko}}, \
  and\ \bibinfo {author} {\bibfnamefont {M.~B.}\ \bibnamefont {Plenio}},\
  }\bibfield  {title} {\enquote {\bibinfo {title} {Robust optical polarization
  of nuclear spin baths using hamiltonian engineering of nitrogen-vacancy
  center quantum dynamics},}\ }\href@noop {} {\bibfield  {journal} {\bibinfo
  {journal} {Sci.\ Adv.}\ }\textbf {\bibinfo {volume} {4}},\ \bibinfo {pages}
  {eaat8978} (\bibinfo {year} {2018})}\BibitemShut {NoStop}%
\bibitem [{\citenamefont {Scheuer}\ \emph {et~al.}(2017)\citenamefont
  {Scheuer}, \citenamefont {Schwartz}, \citenamefont {M{\"u}llar},
  \citenamefont {Chen}, \citenamefont {Dhand}, \citenamefont {Plenio},
  \citenamefont {Naydenov},\ and\ \citenamefont {Jelezko}}]{SSM+17}%
  \BibitemOpen
  \bibfield  {author} {\bibinfo {author} {\bibfnamefont {J.}~\bibnamefont
  {Scheuer}}, \bibinfo {author} {\bibfnamefont {I.}~\bibnamefont {Schwartz}},
  \bibinfo {author} {\bibfnamefont {S.}~\bibnamefont {M{\"u}llar}}, \bibinfo
  {author} {\bibfnamefont {Q.}~\bibnamefont {Chen}}, \bibinfo {author}
  {\bibfnamefont {I.}~\bibnamefont {Dhand}}, \bibinfo {author} {\bibfnamefont
  {M.~B.}\ \bibnamefont {Plenio}}, \bibinfo {author} {\bibfnamefont
  {B.}~\bibnamefont {Naydenov}}, \ and\ \bibinfo {author} {\bibfnamefont
  {F.}~\bibnamefont {Jelezko}},\ }\bibfield  {title} {\enquote {\bibinfo
  {title} {Robust techniques for polarization and detection of nuclear spin
  ensembles},}\ }\href@noop {} {\bibfield  {journal} {\bibinfo  {journal}
  {Phys.\ Rev.\ B}\ }\textbf {\bibinfo {volume} {96}},\ \bibinfo {pages}
  {174436} (\bibinfo {year} {2017})}\BibitemShut {NoStop}%
\bibitem [{\citenamefont {Schmitt}\ \emph {et~al.}(2017)\citenamefont
  {Schmitt}, \citenamefont {Gefen}, \citenamefont {St{\"u}rner}, \citenamefont
  {Unden}, \citenamefont {Wolff}, \citenamefont {M{\"u}ller}, \citenamefont
  {Scheuer}, \citenamefont {Naydenov}, \citenamefont {Markham}, \citenamefont
  {Pezzagna}, \citenamefont {Meijer}, \citenamefont {Schwarz}, \citenamefont
  {Plenio}, \citenamefont {Retzker}, \citenamefont {McGuinness},\ and\
  \citenamefont {Jelezko}}]{SGS+17}%
  \BibitemOpen
  \bibfield  {author} {\bibinfo {author} {\bibfnamefont {S.}~\bibnamefont
  {Schmitt}}, \bibinfo {author} {\bibfnamefont {T.}~\bibnamefont {Gefen}},
  \bibinfo {author} {\bibfnamefont {F.~M.}\ \bibnamefont {St{\"u}rner}},
  \bibinfo {author} {\bibfnamefont {T.}~\bibnamefont {Unden}}, \bibinfo
  {author} {\bibfnamefont {G.}~\bibnamefont {Wolff}}, \bibinfo {author}
  {\bibfnamefont {C.}~\bibnamefont {M{\"u}ller}}, \bibinfo {author}
  {\bibfnamefont {J.}~\bibnamefont {Scheuer}}, \bibinfo {author} {\bibfnamefont
  {B.}~\bibnamefont {Naydenov}}, \bibinfo {author} {\bibfnamefont
  {M.}~\bibnamefont {Markham}}, \bibinfo {author} {\bibfnamefont
  {S.}~\bibnamefont {Pezzagna}}, \bibinfo {author} {\bibfnamefont
  {J.}~\bibnamefont {Meijer}}, \bibinfo {author} {\bibfnamefont
  {I.}~\bibnamefont {Schwarz}}, \bibinfo {author} {\bibfnamefont
  {M.}~\bibnamefont {Plenio}}, \bibinfo {author} {\bibfnamefont
  {A.}~\bibnamefont {Retzker}}, \bibinfo {author} {\bibfnamefont {L.~P.}\
  \bibnamefont {McGuinness}}, \ and\ \bibinfo {author} {\bibfnamefont
  {F.}~\bibnamefont {Jelezko}},\ }\bibfield  {title} {\enquote {\bibinfo
  {title} {Submillihertz magnetic spectroscopy performed with a nanoscale
  quantum sensor},}\ }\href@noop {} {\bibfield  {journal} {\bibinfo  {journal}
  {Science}\ }\textbf {\bibinfo {volume} {356}},\ \bibinfo {pages} {832}
  (\bibinfo {year} {2017})}\BibitemShut {NoStop}%
\bibitem [{\citenamefont {Boss}\ \emph {et~al.}(2017)\citenamefont {Boss},
  \citenamefont {Cujia}, \citenamefont {Zopes},\ and\ \citenamefont
  {Degen}}]{BCZD17}%
  \BibitemOpen
  \bibfield  {author} {\bibinfo {author} {\bibfnamefont {J.~M.}\ \bibnamefont
  {Boss}}, \bibinfo {author} {\bibfnamefont {K.~S.}\ \bibnamefont {Cujia}},
  \bibinfo {author} {\bibfnamefont {J.}~\bibnamefont {Zopes}}, \ and\ \bibinfo
  {author} {\bibfnamefont {C.~L.}\ \bibnamefont {Degen}},\ }\bibfield  {title}
  {\enquote {\bibinfo {title} {Quantum sensing with arbitrary frequency
  resolusion},}\ }\href@noop {} {\bibfield  {journal} {\bibinfo  {journal}
  {Science}\ }\textbf {\bibinfo {volume} {356}},\ \bibinfo {pages} {837}
  (\bibinfo {year} {2017})}\BibitemShut {NoStop}%
\bibitem [{\citenamefont {Glenn}\ \emph {et~al.}(2018)\citenamefont {Glenn},
  \citenamefont {Bucher}, \citenamefont {Lee}, \citenamefont {Lukin},
  \citenamefont {Park},\ and\ \citenamefont {Walsworth}}]{GBL+18}%
  \BibitemOpen
  \bibfield  {author} {\bibinfo {author} {\bibfnamefont {D.~R.}\ \bibnamefont
  {Glenn}}, \bibinfo {author} {\bibfnamefont {D.~B.}\ \bibnamefont {Bucher}},
  \bibinfo {author} {\bibfnamefont {J.}~\bibnamefont {Lee}}, \bibinfo {author}
  {\bibfnamefont {M.~D.}\ \bibnamefont {Lukin}}, \bibinfo {author}
  {\bibfnamefont {H.}~\bibnamefont {Park}}, \ and\ \bibinfo {author}
  {\bibfnamefont {R.~L.}\ \bibnamefont {Walsworth}},\ }\bibfield  {title}
  {\enquote {\bibinfo {title} {High-resolution magnetic resonance spectroscopy
  using a solid-state spin sensor},}\ }\href@noop {} {\bibfield  {journal}
  {\bibinfo  {journal} {Nature}\ }\textbf {\bibinfo {volume} {555}},\ \bibinfo
  {pages} {351} (\bibinfo {year} {2018})}\BibitemShut {NoStop}%
\bibitem [{\citenamefont {Pfender}\ \emph {et~al.}(2019)\citenamefont
  {Pfender}, \citenamefont {Wang}, \citenamefont {Sumiya}, \citenamefont
  {Onoda}, \citenamefont {Yang}, \citenamefont {Dasari}, \citenamefont
  {Neumann}, \citenamefont {Pan}, \citenamefont {Isoya}, \citenamefont {Liu},\
  and\ \citenamefont {Wrachtrup}}]{PWS+19}%
  \BibitemOpen
  \bibfield  {author} {\bibinfo {author} {\bibfnamefont {M.}~\bibnamefont
  {Pfender}}, \bibinfo {author} {\bibfnamefont {P.}~\bibnamefont {Wang}},
  \bibinfo {author} {\bibfnamefont {H.}~\bibnamefont {Sumiya}}, \bibinfo
  {author} {\bibfnamefont {S.}~\bibnamefont {Onoda}}, \bibinfo {author}
  {\bibfnamefont {W.}~\bibnamefont {Yang}}, \bibinfo {author} {\bibfnamefont
  {D.~B.~R.}\ \bibnamefont {Dasari}}, \bibinfo {author} {\bibfnamefont
  {P.}~\bibnamefont {Neumann}}, \bibinfo {author} {\bibfnamefont {X.-Y.}\
  \bibnamefont {Pan}}, \bibinfo {author} {\bibfnamefont {J.}~\bibnamefont
  {Isoya}}, \bibinfo {author} {\bibfnamefont {R.-B.}\ \bibnamefont {Liu}}, \
  and\ \bibinfo {author} {\bibfnamefont {J.}~\bibnamefont {Wrachtrup}},\
  }\bibfield  {title} {\enquote {\bibinfo {title} {High-resolution spectroscopy
  of single nuclear spins via sequential weak measurements},}\ }\href@noop {}
  {\bibfield  {journal} {\bibinfo  {journal} {Nat.\ Commun.}\ }\textbf
  {\bibinfo {volume} {10}},\ \bibinfo {pages} {594} (\bibinfo {year}
  {2019})}\BibitemShut {NoStop}%
\bibitem [{\citenamefont {Cujia}\ \emph {et~al.}(2019)\citenamefont {Cujia},
  \citenamefont {Boss}, \citenamefont {Herb}, \citenamefont {Zopes},\ and\
  \citenamefont {Degen}}]{CBH+19}%
  \BibitemOpen
  \bibfield  {author} {\bibinfo {author} {\bibfnamefont {K.~S.}\ \bibnamefont
  {Cujia}}, \bibinfo {author} {\bibfnamefont {J.~M.}\ \bibnamefont {Boss}},
  \bibinfo {author} {\bibfnamefont {K.}~\bibnamefont {Herb}}, \bibinfo {author}
  {\bibfnamefont {J.}~\bibnamefont {Zopes}}, \ and\ \bibinfo {author}
  {\bibfnamefont {C.~L.}\ \bibnamefont {Degen}},\ }\bibfield  {title} {\enquote
  {\bibinfo {title} {Tracking the precession of single nuclear spins by weak
  measurements},}\ }\href@noop {} {\bibfield  {journal} {\bibinfo  {journal}
  {Nature}\ }\textbf {\bibinfo {volume} {571}},\ \bibinfo {pages} {230}
  (\bibinfo {year} {2019})}\BibitemShut {NoStop}%
\bibitem [{\citenamefont {Pham}\ \emph {et~al.}(2016)\citenamefont {Pham},
  \citenamefont {DeVience}, \citenamefont {Casola}, \citenamefont {Lovchinsky},
  \citenamefont {Sushkov}, \citenamefont {Bersin}, \citenamefont {Lee},
  \citenamefont {Urbach}, \citenamefont {Cappellaro}, \citenamefont {Park},
  \citenamefont {Yacoby}, \citenamefont {Lukin},\ and\ \citenamefont
  {Walsworth}}]{PDC+16}%
  \BibitemOpen
  \bibfield  {author} {\bibinfo {author} {\bibfnamefont {L.~M.}\ \bibnamefont
  {Pham}}, \bibinfo {author} {\bibfnamefont {S.~J.}\ \bibnamefont {DeVience}},
  \bibinfo {author} {\bibfnamefont {F.}~\bibnamefont {Casola}}, \bibinfo
  {author} {\bibfnamefont {I.}~\bibnamefont {Lovchinsky}}, \bibinfo {author}
  {\bibfnamefont {A.~O.}\ \bibnamefont {Sushkov}}, \bibinfo {author}
  {\bibfnamefont {E.}~\bibnamefont {Bersin}}, \bibinfo {author} {\bibfnamefont
  {J.}~\bibnamefont {Lee}}, \bibinfo {author} {\bibfnamefont {E.}~\bibnamefont
  {Urbach}}, \bibinfo {author} {\bibfnamefont {P.}~\bibnamefont {Cappellaro}},
  \bibinfo {author} {\bibfnamefont {H.}~\bibnamefont {Park}}, \bibinfo {author}
  {\bibfnamefont {A.}~\bibnamefont {Yacoby}}, \bibinfo {author} {\bibfnamefont
  {M.}~\bibnamefont {Lukin}}, \ and\ \bibinfo {author} {\bibfnamefont {R.~L.}\
  \bibnamefont {Walsworth}},\ }\bibfield  {title} {\enquote {\bibinfo {title}
  {{NMR} technique for determining the depth of shallow nitrogen-vacancy
  centers in diamond},}\ }\href@noop {} {\bibfield  {journal} {\bibinfo
  {journal} {Phys.\ Rev.\ B}\ }\textbf {\bibinfo {volume} {93}},\ \bibinfo
  {pages} {045425} (\bibinfo {year} {2016})}\BibitemShut {NoStop}%
\bibitem [{\citenamefont {Glover}\ \emph {et~al.}(2003)\citenamefont {Glover},
  \citenamefont {Newton}, \citenamefont {Martineau}, \citenamefont {Twitchen},\
  and\ \citenamefont {Baker}}]{GNM+03}%
  \BibitemOpen
  \bibfield  {author} {\bibinfo {author} {\bibfnamefont {C.}~\bibnamefont
  {Glover}}, \bibinfo {author} {\bibfnamefont {M.~E.}\ \bibnamefont {Newton}},
  \bibinfo {author} {\bibfnamefont {P.}~\bibnamefont {Martineau}}, \bibinfo
  {author} {\bibfnamefont {D.~J.}\ \bibnamefont {Twitchen}}, \ and\ \bibinfo
  {author} {\bibfnamefont {J.}~\bibnamefont {Baker}},\ }\bibfield  {title}
  {\enquote {\bibinfo {title} {Hydrogen {I}ncorporation in {D}iamond: {T}he
  {N}itrogen-{V}acancy-{H}ydrogen {C}omplex},}\ }\href@noop {} {\bibfield
  {journal} {\bibinfo  {journal} {Phys.\ Rev.\ Lett.}\ }\textbf {\bibinfo
  {volume} {90}},\ \bibinfo {pages} {185507} (\bibinfo {year}
  {2003})}\BibitemShut {NoStop}%
\bibitem [{\citenamefont {Goss}\ \emph {et~al.}(2003)\citenamefont {Goss},
  \citenamefont {Briddon}, \citenamefont {Jones},\ and\ \citenamefont
  {Sque}}]{GBJS03}%
  \BibitemOpen
  \bibfield  {author} {\bibinfo {author} {\bibfnamefont {J.~P.}\ \bibnamefont
  {Goss}}, \bibinfo {author} {\bibfnamefont {P.~R.}\ \bibnamefont {Briddon}},
  \bibinfo {author} {\bibfnamefont {R.}~\bibnamefont {Jones}}, \ and\ \bibinfo
  {author} {\bibfnamefont {S.}~\bibnamefont {Sque}},\ }\bibfield  {title}
  {\enquote {\bibinfo {title} {The vacancy--nitrogen--hydrogen complex in
  diamond: a potential deep centre in chemical vapour deposited material},}\
  }\href@noop {} {\bibfield  {journal} {\bibinfo  {journal} {J.\ Phys.:
  Condens.\ Matter}\ }\textbf {\bibinfo {volume} {15}},\ \bibinfo {pages}
  {S2903} (\bibinfo {year} {2003})}\BibitemShut {NoStop}%
\bibitem [{\citenamefont {Edmonds}\ \emph {et~al.}(2012)\citenamefont
  {Edmonds}, \citenamefont {D'Haenens-Johansson}, \citenamefont {Cruddace},
  \citenamefont {Newton}, \citenamefont {Fu}, \citenamefont {Santori},
  \citenamefont {Beausoleil}, \citenamefont {Twitchen},\ and\ \citenamefont
  {Markham}}]{EDC+12}%
  \BibitemOpen
  \bibfield  {author} {\bibinfo {author} {\bibfnamefont {A.~M.}\ \bibnamefont
  {Edmonds}}, \bibinfo {author} {\bibfnamefont {U.~F.~S.}\ \bibnamefont
  {D'Haenens-Johansson}}, \bibinfo {author} {\bibfnamefont {R.~J.}\
  \bibnamefont {Cruddace}}, \bibinfo {author} {\bibfnamefont {M.~E.}\
  \bibnamefont {Newton}}, \bibinfo {author} {\bibfnamefont {K.-M.~C.}\
  \bibnamefont {Fu}}, \bibinfo {author} {\bibfnamefont {C.}~\bibnamefont
  {Santori}}, \bibinfo {author} {\bibfnamefont {R.~G.}\ \bibnamefont
  {Beausoleil}}, \bibinfo {author} {\bibfnamefont {D.~J.}\ \bibnamefont
  {Twitchen}}, \ and\ \bibinfo {author} {\bibfnamefont {M.~L.}\ \bibnamefont
  {Markham}},\ }\bibfield  {title} {\enquote {\bibinfo {title} {Production of
  oriented nitrogen-vacancy color centers in synthetic diamond},}\ }\href@noop
  {} {\bibfield  {journal} {\bibinfo  {journal} {Rev.\ Mod.\ Phys.}\ }\textbf
  {\bibinfo {volume} {86}},\ \bibinfo {pages} {, 035201} (\bibinfo {year}
  {2012})}\BibitemShut {NoStop}%
\bibitem [{\citenamefont {Degen}, \citenamefont {Reinhard},\ and\ \citenamefont
  {Cappellaro}(2017)}]{DRC17}%
  \BibitemOpen
  \bibfield  {author} {\bibinfo {author} {\bibfnamefont {C.~L.}\ \bibnamefont
  {Degen}}, \bibinfo {author} {\bibfnamefont {F.}~\bibnamefont {Reinhard}}, \
  and\ \bibinfo {author} {\bibfnamefont {P.}~\bibnamefont {Cappellaro}},\
  }\bibfield  {title} {\enquote {\bibinfo {title} {Quantum sensing},}\
  }\href@noop {} {\bibfield  {journal} {\bibinfo  {journal} {Rev.\ Mod.\
  Phys.}\ }\textbf {\bibinfo {volume} {89}},\ \bibinfo {pages} {035002}
  (\bibinfo {year} {2017})}\BibitemShut {NoStop}%
\end{thebibliography}%
\end{document}